\documentclass[conference]{IEEEtran}
\IEEEoverridecommandlockouts
\usepackage{cite}
\usepackage{amsmath,amssymb,amsfonts}
\usepackage{algorithmic}
\usepackage{graphicx}
\usepackage{textcomp}
\usepackage{xcolor}
\usepackage{soul}
\def\BibTeX{{\rm B\kern-.05em{\sc i\kern-.025em b}\kern-.08em
    T\kern-.1667em\lower.7ex\hbox{E}\kern-.125emX}}

\begin{document}

\title{Applying the Ego Network Model to Cross-Target Stance Detection}

\author{
    \IEEEauthorblockN{
        Jack Tacchi\IEEEauthorrefmark{1}\IEEEauthorrefmark{3}\IEEEauthorrefmark{4}, 
        Parisa Jamadi Khiabani\IEEEauthorrefmark{2}\IEEEauthorrefmark{4}, 
        Arkaitz Zubiaga\IEEEauthorrefmark{2}, 
        Chiara Boldrini\IEEEauthorrefmark{1}, 
        Andrea Passarella\IEEEauthorrefmark{1}
    }
    \IEEEauthorblockA{\IEEEauthorrefmark{1} \textit{Istituto di Informatica e Telematica}, \textit{Consiglio Nazionale delle Ricerche}, Pisa, Italy}
    \IEEEauthorblockA{\IEEEauthorrefmark{2} \textit{School of Electronic Engineering and 
    Computer Science}, \textit{Queen Mary University}, London, United Kingdom}
    \IEEEauthorblockA{\IEEEauthorrefmark{3} \textit{Scuola Normale Superiore}, Pisa, Italy\vspace{-1.6em}}
}

\maketitle

\def\thefootnote{§}\footnotetext{Co-first authors with equal contribution and importance.}
\def\thefootnote{\arabic{footnote}}

\begin{abstract}
Understanding human interactions and social structures is an incredibly important task, especially in such an interconnected world. One task that facilitates this is Stance Detection, which predicts the opinion or attitude of a text towards a target entity. Traditionally, this has often been done mainly via the use of text-based approaches, however, recent work has produced a model (CT-TN) that leverages information about a user's social network to help predict their stance, outperforming certain cross-target text-based approaches. Unfortunately, the data required for such graph-based approaches is not always available. This paper proposes two novel tools for Stance Detection: the Ego Network Model (ENM) and the Signed Ego Network Model (SENM). These models are founded in anthropological and psychological studies and have been used within the context of social network analysis and related tasks (e.g., link prediction). Stance Detection predictions obtained using these features achieve a level of accuracy similar to the graph-based features used by CT-TN while requiring less and more easily obtainable data. In addition to this, the performances of the inner and outer circles of the ENM, representing stronger and weaker social ties, respectively are compared. Surprisingly, the outer circles, which contain more numerous but less intimate connections, are more useful for predicting stance.
\end{abstract}


\vspace{-1em}
\section{Introduction}
\label{sec:introduction}

Humans have always been social animals. 
Everyday we interact countless times with one another and these interactions form the basis of our modern societies.
It has even been argued that our ability to maintain larger social groups was the primary reason that the volume of our brains increased significantly in size~\cite{Dunbar_1998}.
What's more, since the advent of the internet, humans have also been able to interact with each other regardless of geographical location.
Because humans are now more interconnected than ever, understanding social connections and how they contribute to the spreading of ideas and opinions has never been more important.

The exponential expansion of social networks has introduced fresh obstacles in the realm of information retrieval ~\cite{Jamadi_2020}. 
One research task that deals directly with opinions on social networks is Stance Detection (SD), which aims to predict the stance of a given text towards a target entity~\cite{Biber_1988}. 
Of course, being able to monitor the opinions of individuals or even overall trends in larger communities and populations can be extremely impactful.
Especially, given its application to politics, where it can be used to quickly understand how people feel towards a given topic or to predict how they will vote. 

\begin{figure}[h]
  \vspace{-1em}
  \centering
  \includegraphics[width=0.34\textwidth,height=0.2\textheight]{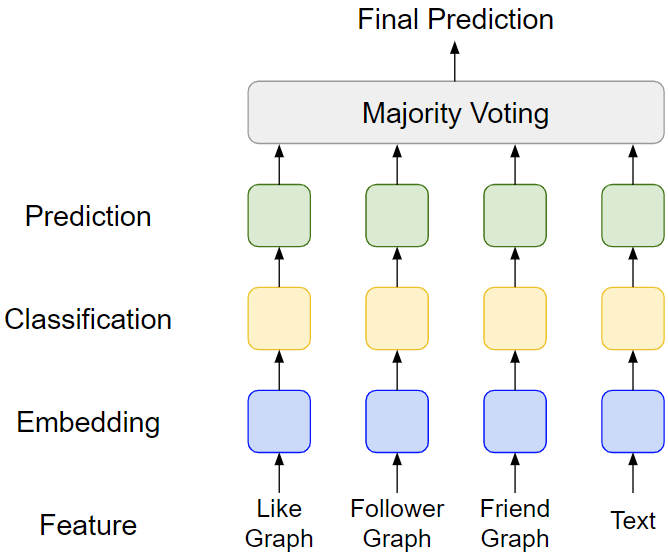}
  \caption{Architecture of the CT-TN model.}
  \label{fig:CT-TN_model}
  \vspace{-0.5em}
\end{figure}

SD has often viewed interactions in isolation, predicting a user's opinion towards a given entity purely based on what they have written (e.g.~\cite{Wei_2019}).
However, recent research has shown that considering an individual's surrounding social network can greatly improve the accuracy of SD~\cite{Khiabani_2023}, highlighting the influence of social connections on opinions.
Specifically, a model called CT-TN (Cross-Target Text-Net), which uses predictions from a more traditional text-based model, RoBERTa~\cite{Liu_2019}, together with multiple network features from the X (formerly Twitter) social media platform: likes (a list of users whose posts have been liked by the target user), followers (the users who follow the target user) and friends (the users who are followed by the target user).
This was done by creating an embedding for each feature and passing it through a classification model to obtain a prediction for each feature.
The final prediction was then generated based on a majority vote of all the features (see Figure~\ref{fig:CT-TN_model}).
The CT-TN model outperformed other competitive models, such as CrossNET~\cite{Xu_2018} and TGA-Net~\cite{Allaway_2020} in six different experimental conditions.

However, the different network features that CT-TN requires are not always available.
Indeed, the multiple different data sources required for CT-TN may be impossible or extremely costly to obtain in many situations.
Therefore, it would be pertinent to investigate alternative approaches or features that are more parsimonious.

To this effect, in this work, we exploit models of users' personal networks, grounded on well-established findings in anthropology. In addition to providing solid quantitative models of the structure of users' social relationships, which can be obtained whenever users communicate publicly, thus minimising restrictions from the aforementioned shortcomings as much as possible.
Indeed, given the aforementioned importance of understanding how humans communicate, it is unsurprising that this topic had been researched long before the internet.
For instance, anthropological and psychological studies have found that the number of relationships that an individual is able to maintain is remarkably consistent across members of the same species~\cite{Dunbar_1992}.
The maximum maintainable group size is correlated with the proportional size of the neocortex part of said species' brain; strongly suggesting that a group's size is innately limited by the cognitive ability of the species within it~\cite{Dunbar_1998}.
What's more, these connections can be sorted by contact frequency into a series of concentric groups~\cite{Dunbar_1995}, with a smaller number of high-frequency connections at the centre and larger numbers of less-frequent connections towards the edges.
\vspace{-0.2em}

\begin{figure}[h]
  \centering
  \includegraphics[scale=0.29]{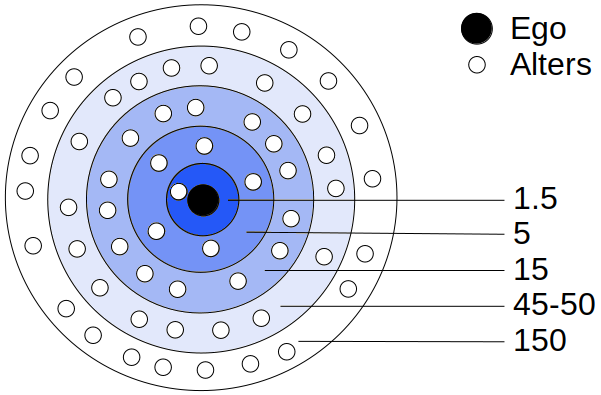}
  \caption{The Ego Network Model, with the expected number of Alters of each circle (for humans).}
  \label{fig:ego_network_model}
  \vspace{-1.5em}
\end{figure}

A representation of this structure, known as the Ego Network Model (ENM), places a target individual (the Ego, from which the model takes its name) at the centre and surrounds them by all of their connections (Alters) (see Figure~\ref{fig:ego_network_model}).
For humans, the expected sizes of the concentric groups range from 5 (support clique), to 15 (sympathy group), then around 45-50 (affinity group) and finally 150 (active network)~\cite{Dunbar_1995}. 
An additional inner circle of 1-2 connections has also been found in many online contexts~\cite{Arnaboldi_2013a}.
Interestingly, these groups increase in size by a factor of around 3 each time, and this scaling factor has also been found for certain non-human animals, such as other primates and even birds~\cite{Dunbar_2014}.
\vspace{-0.2em}

The ENM has been studied not only offline but has also been used to reveal many novel insights about communications in online contexts~\cite{Boldrini_2018}.
For instance, the ENM has been used to discover differences between the behaviours of various types of users, including journalists~\cite{Toprak_2021} and members of different online communities~\cite{Tacchi_2023}. 
It has also been shown that using ENM information can significantly improve ``classical" tasks in Online Social Networks (OSNs), such as link prediction~\cite{Toprak_2023}. 
Moreover, recent extensions of ENM have been proposed to label links in the Ego Network with positive or negative signs, depending on whether the corresponding relationship between Ego and Alter displays a predominantly positive or negative sentiment~\cite{Tacchi_2022}. 
This extension, called the Signed ENM (SENM) has been used to characterise, in detail, the prevailing polarity of social relationships in OSNs, in contrast to the offline world~\cite{Tacchi_2024}.
Therefore, the ENM and SENM could reasonably be expected to provide significant information pertinent to the task of SD.
Indeed, Ego Networks go beyond providing mere structural information but can contribute semantically to a better knowledge of the underlying mechanisms of a network and, therefore, provide pertinent contextual insights for SD.
In addition, the base ENM requires only a list of Ego-Alter pairs and the frequencies of their interaction; data which can be easily obtained just by monitoring public interactions.
\vspace{-0.2em}

\subsection{Contributions}
\vspace{-0.2em}

Given the potential restrictions of data availability discussed above, it is important to have a toolbox of diverse approaches to ensure that researchers have a viable means of tackling as many situations as possible.
Indeed, the authors of the CT-TN model specifically advocate for exploring further network features to enhance SD~\cite{Khiabani_2023}.
In response to this, the current work proposes two features from adjacent network research areas for use within the task of SD: the ENM and the SENM.
Furthermore, this paper demonstrates that these features perform better than a text-only approach, RoBERTa, and similarly, although slightly worse, than the cutting-edge CT-TN model.
Thus, demonstrating that these novel features are viable alternatives for SD when not all the required features for CT-TN can be obtained.
Finally, a comparison of the performances between the inner and outer circles provides evidence that a user's less intimate but more numerous outer connections have more of an impact on their stance.

\section{Background}

\subsection{Stance Detection}
\label{sec:stance_detection}

The focus of most SD studies is to classify target dependency in one of four main ways: Target-Specific, Multi-Related Targets, Target-Independent and Cross-Target~\cite{Alturayeif_2023}. 
For this paper, we primarily concentrate on Cross-Target Stance Detection (CTSD), which is when a model is trained using data for one target entity (source) and then tested on a different, although related, target entity (destination). 
For example, a model trained using texts containing opinions towards Joe Biden could be used to predict the stance of texts concerning another politician, such as Donald Trump or Bernie Sanders.

Approaches to CTSD, as well as to SD more generally, differ based on the text's context and the particular relationship being discussed. 
On social media platforms such as X, there's often a focus on discerning the author's stance (supportive, opposing, or neutral) towards a specific proposition or target~\cite{Mohammad_2016}. 
Recent advances in SD encompass a range of linguistic features, such as word or character n-grams, dependency parse trees, and lexicons~\cite{Sun_2017}. 
Moreover, there has been a shift towards end-to-end neural network methods that independently learn topics and opinions, integrating them via mechanisms such as memory networks or bidirectional conditional Long Short-Term Memory models (LSTMs)~\cite{Augenstein_2016}.

Past studies have primarily relied on the text of a post to gauge its stance, neglecting the valuable insights that other features within social media platforms could offer. 
However, the performance of the aforementioned CT-TN model demonstrates the importance of considering structural features of the surrounding social network.
Thus, knowledge about a user's connections can reveal important insights about the user themself and, therefore, about the texts they author.

\subsection{The Ego Network Model}
\label{sec:ego_network_model}

As mentioned in Section~\ref{sec:introduction}, the ENM views a social network from the point of view of an individual Ego and organises their Alters around them based on their contact frequency, while the SENM extends this by adding the addition of signed relationships.
Specifically, these signed relationships are obtained by analysing the sentiment of the interactions between each Ego and Alter, thus obtaining a list of sentiments for each relationship, which can then be used to infer an overall sentiment for the relationship as a whole.
It has been observed that many different types of relationships start to have negative effects on the people involved once the ratio of negative interactions generated by that relationship passes a certain ratio: around 17\%~\cite{Gottman_1995}.
Therefore, this threshold is used to infer the sentiments of relationships based on their interactions.

While the Signed Ego Network provides an additional layer of information, it does require text for each interaction in a dataset.
By contrast, the unsigned Ego Network requires only the frequency of interactions between each pair of users, without the need for any text.

\section{Methodology}
\label{sec:methodology}
\vspace{-0.2em}

\subsection{Performing Cross-Target Stance Detection}
\label{sec:performing_stance_Detection}

\subsubsection{Feature Embeddings}
In order to compute SD predictions, the data first need to be transformed into representations that are readable by a prediction model.
For this, node2Vec~\cite{Grover_2016} was applied to each of the previously established graph-based features (likes, followers, friends) as well as the novel Ego Networks and Signed Ego Networks, which, although they are converted into the same vector-space representation, can be better thought of as proxy measures of the way humans function socially.
Node2vec is an unsupervised Deep Learning algorithm that uses a flexible, biased, random walk procedure to explore networks.
The visited nodes can then be transformed into a vector space representation using a variety of methods, such as skip-grams or a continuous bag-of-words~\cite{Grover_2016}.
This is similar to how the word2vec algorithm~\cite{Mikolov_2013} treats words (nodes) and sentences (walks).

In addition to the graph-based features, text-based predictions were also used.
These were generated using RoBERTa~\cite{Mikolov_2013}, an incredibly well-performing pretrained model that is used for many different natural language processing tasks.
RoBERTa maps every token in a sentence to a vector representation in a continuous space.

As mentioned in Section~\ref{sec:introduction}, the CT-TN model takes the predictions of each of these features, RoBERTa, likes, followers and friends, and obtains a final prediction using majority voting, where each feature's prediction acts as a vote for either ``FAVOR'' or ``AGAINST''.
This allows for a thorough analysis of both textual and social network information, providing valuable insights for CTSD.

Aside from the aforementioned features, this paper also investigates two novel graph-based features: Ego Networks and Signed Ego Networks.
These are also converted to a vector space representation using node2vec.
Further details on how the Ego Networks are obtained are explained in Section~\ref{sec:computing_ego_networks}.



\subsubsection{Model Hyperparameters}
\label{sec:Model_Hyperparameters}
Each of the features was used to train a neural network model with two hidden layers for classification task.
For the text-based embedding, this was done using RoBERTa, a batch size of 128, a dropout of 0.2, a learning rate of 3e-5 (AdamW), and 40 epochs. 
For the graph-based embeddings, this was done using node2vec, with the same batch size and dropout, a learning rate of 1e-2 (SGD), and 100 epochs.

\subsubsection{Experimental Settings}
\label{sec:Experiment_Settings}

The CT-TN model and the individual RoBERTa feature predictions were used as baselines against which to test the two Ego Network features.
These were all prepared using few-shot cross-target training, whereby the training data consisted of roughly 1,000 source target data points with 4 injections of destination target texts, increasing in size by increments of 100, from 100-shot to 400-shot (inclusive). 
For example, the Biden-Trump predictions were obtained by training on around 1,000 Biden texts with 100 Trump texts for the 100-shot condition, with 200 Trump tweets for the 200-shot condition, and so on. 
The stance predictions were then tested using between 500 and 800 data points (depending on the amount of remaining unseen data) that were solely related to the destination target (i.e. only using Trump-related texts for the aforementioned example).
We conducted these experiments with five different random seeds: 24, 524, 1024, 1524, and 2024. 
Finally, we averaged the results from these five seeds for each shot size.

\subsection{Computing Ego Networks}
\label{sec:computing_ego_networks}



\subsubsection{Computing Ego Networks}
An Ego Network can be obtained for each active\footnote{An active user is defined as a user, with a timeline of at least 6 months, that posts at least once every three days for half of the months that they are included in the data, based on previous research~\cite{Boldrini_2018}} user in the data by calculating the interaction frequencies of each of their relationships and then applying a clustering algorithm to them.
One of the most popular methods for this is MeanShift~\cite{Fukunaga_1975}, an unsupervised algorithm that automatically finds the most appropriate number of circles for an Ego~\cite{Boldrini_2018}, and so that is the algorithm that was employed for this work.

Additionally, the unsigned Ego Network feature was also separated into inner circles (1 and 2) and outer circles (3+), to better understand the importance of the different levels of the ENM for SD.

\subsubsection{Generating Signed Connections}
Once the unsigned Ego Networks have been obtained, it is relatively simple to generate the signed version.
The interactions were grouped for each Ego-Alter pair and then a model, called Valence Aware Dictionary and sEntiment Reasoner (VADER)~\cite{Hutto_2014}, was used to obtain a sentiment for each interaction.
VADER is a very competitive sentiment analysis model that is specifically designed for performing sentiment analysis on English-language tweets.
It is also the model used in the original SENM research paper~\cite{Tacchi_2022}.

Once sentiments had been obtained for the individual interactions, the psychology-based threshold of 17\%~\cite{Gottman_1995}, mentioned in Section~\ref{sec:ego_network_model}, was used to obtain a sentiment label for each relationship, resulting in signed Ego Networks.

\subsection{Data}
\label{sec:data}
The data used in this study come from a well-established and publicly available set: P-Stance~\cite{Li_2021}.
This dataset contains 21,574 English-language tweets collected during the 2020 U.S. presidential election.
These tweets were specifically collected to be used for SD and each one is associated with one of 3 targets, Joe Biden, Donald Trump or Bernie Sanders, and a corresponding stance label, either ``AGAINST'' or ``FAVOR''. 
Hashtags, such as ``\#BidenForPresident'' and ``\#NeverBernie'' were used to both search for the tweets and to determine their target and stance.
While the original dataset only included the text and stance of each tweet, the authors of the P-Stance dataset provided 9,307 tweet IDs upon request, allowing further data to be collected for each tweet, including information about the authoring users.
In addition to providing the information required for computing the users' Ego Networks (see Section~\ref{sec:computing_ego_networks}), this also made it possible to obtain, for each user, the remaining features required by CT-TN: likes, followers and friends.

\vspace{-0.3em}

\section{Results}
\label{sec:results}

\begin{figure*}[h]
  \centering
  \includegraphics[scale=0.55]{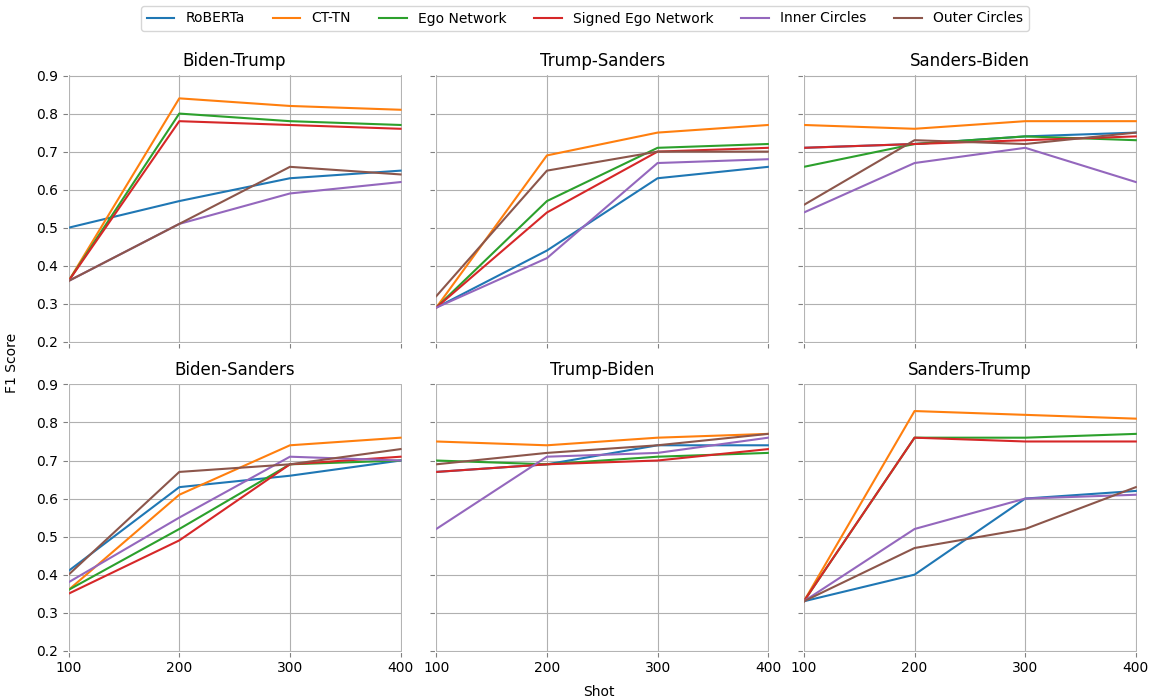}
  \vspace{-0.8em}
  \caption{Graphs displaying the performances of the individual text-based RoBERTa, the CT-TN model and the signed and unsigned Ego Networks features, as well as the unsigned inner and outer circle features. The shot number is displayed along the X axis and the averaged macro F1 scores along the Y axis.}
  \label{fig:ctsd_results}
  \vspace{-1.5em}
\end{figure*}

The performances of the CT-TN model, RoBERTa, and the two Ego Network features can be seen in Figure~\ref{fig:ctsd_results}. 
Overall, they all perform very well, with most reaching a macro F1 score of above 0.7 before 400-shot for all target pairs, with CT-TN sometimes even going above 0.8.
However, RoBERTa does not perform quite as well as the others, and only achieves macro F1 scores of around mid-0.6 for half of the target pairs.

Surprisingly, the signed and unsigned Ego Networks' F1 scores are very close, being within 0.01 of each other for 5 of the 6 target pairs (at 400-shot), and within 0.02 for the sixth pair (Sanders-Trump).
This suggests that the additional information of signed connections does not provide a significant amount of information for the task of CTSD.
Rather, it appears that the people we interact with regularly have an impact on our stances regardless of whether we have a negative or positive relationship with them.

Next, observing the outer circles, one can see that they perform similarly to, and often even outperform, the full Ego Network.
However, they are less consistent, as displayed by the Biden-Trump and Sanders-Trump target pairs.
By comparison, the inner circles perform slightly worse overall, with performances closer to those of RoBERTa. 
Since the outer circles contain weaker social relationships, it seems that weaker, but more numerous, ties are more informative than stronger, but less numerous, ones when it comes to stance prediction. 
This is rather surprising given that previous research on a similar network-based task, link prediction, found that the more intimate inner circles are better predictors of where new relations will form~\cite{Toprak_2023}.
Thus, there seems to be a disconnect between how we form new connections and how we are influenced by them.
Indeed, paired with the fact that the signed and unsigned ENMs performed similarly, it appears that the existence of a social connection may influence an individual regardless of any qualitative aspects, such as closeness or polarity.

The Ego Networks appear to perform slightly worse than the CT-TN model.
However, as they only require interaction data, they could be used as a viable alternative whenever specific network features are not provided or obtainable for a given dataset. 
Moreover, as the signed and unsigned Ego Networks achieved similar performances, one could focus on employing the unsigned version, which would require even less data: only the frequencies of interactions, without the need for their texts.
\vspace{-1.5em}

\section{Conclusion}
\label{sec:conclusion}

This study has highlighted potential limitations with previous approaches to CTSD (and SD more generally) due to the availability, or rather the unavailability, of data.
This is especially relevant given recent restrictions to data accessibility on X, most notably the discontinuation of the Academic API, which has been one of the largest and most popular social network data sources for academic use for over a decade~\cite{Arnaboldi_2013a,Boldrini_2018,Tacchi_2022,Tacchi_2024}.

Addressing this limitation, this paper has proposed two novel graph-based features that can be used for SD: the ENM and the SENM.
The latter requires only text-based interactions between users and the former only requires interaction frequencies, meaning that the content of the interactions can remain hidden.
These features perform consistently well for CTSD (achieving macro F1 scores of at least 0.71, after 400 shots, for all 6 target pairs tested in this paper).
Their performances are only slightly worse than a previously established and cutting-edge model, CT-TN.
Ego Networks present viable alternatives that could be used when not all the features required for CT-TN (or similar models) are obtainable.

Finally, by observing the different performances between the inner and outer circles of the ENM, it appears that, while the inclusion of both leads to a more consistent performance across different target pairs, the outer circles on their own can often perform just as well and sometimes better.
By contrast, the inner circles do not perform as well, suggesting that the greater number of less intimate connections in the outer circles are more important for predicting a user's stance.
\vspace{-0.5em}

\bibliographystyle{IEEEtran}
\bibliography{biblio}

\end{document}